\documentclass[12pt]{article}

\usepackage{graphicx}

\begin{document}

\centerline {\Large  A Study of Elementary Excitations of Liquid Helium-4}

\smallskip
\centerline {\Large   Using Macro-orbital Microscopic Theory}

\vspace{0.5cm} 

\centerline {\bf Yatendra S. Jain}

\vspace{.3cm} 

\centerline {Department of Physics, North-Eastern Hill University,} 

\centerline {Shillong-793 022, Meghalaya, India}

\begin{abstract}  
Energy of elementary excitations [$E(Q)$] and the anomalous nature 
of small $Q$ phonons in $He-II$ are studied by using our 
macro-orbital microscopic theory of a system of interacting bosons 
({\it Cond-mat/0606571}).  It is observed that: (i) the experimental 
$E(Q) = E(Q)_{exp}$ of $He-II$ not only agrees with our theoretical 
relation, $E(Q) = \hbar^2Q^2/4mS(Q)$, but also supports an important 
conclusion of Price that $S(0)$ should have zero value for quantum 
fluids, and (ii) Feynman's energy of excitations $E(Q)_{Fyn} = 
\hbar^2Q^2/2mS(Q)$ equals $\approx 2E(Q)_{exp}$ even at low $Q$. 
Three problems with the Feynman's inference that $E(Q)_{Fyn}$ has good 
agreement with $E(Q)_{Exp}$ at low $Q$ are identified. It is 
argued that the theory can also be used to understand similar spectrum of 
the BEC state of a dilute gas reported by O'Dell {\it et. al.}    

\end{abstract}   

{\it email} :  ysjain@email.com



\bigskip

\centerline{\bf 1. Introduction}

\bigskip
Elementary excitations of {\it Liquid} $^4He$, -a typical 
{\it system of interacting bosons} (SIB), have been a subject 
of extensive theoretical and experimental investigation which 
have been elegantly reviewed in [1].  While experimental 
studies, performed under different physical conditions, are 
reported in [2-9], development of their microscopic understanding 
started with Feynman [10] who obtained 
$$E(Q) = E(Q)_{Fyn} = \frac{\hbar^2Q^2}{2mS(Q)}   \eqno(1)$$

\noindent
where $E(Q)$ and $S(Q)$, respectively, represent the energy of 
excitations and structure factor of the system at wave vector $Q$. 
Guided by the observation, that $E(Q)_{Fyn}$ has close 
agreement with experimental $E(Q) [= E(Q)_{exp}]$ of $He-II$ at low 
$Q$ and equals $\approx 2E(Q)_{exp}$ at high $Q$, Feynman and 
Cohen [11] used the phenomenon of back flow to 
obtain a relation of better agreement with $E(Q)_{exp}$ but the 
difference at higher $Q$ could not be reduced to desired level.  
Consequently, several researchers [12-14] used different 
mathematical tools and computational techniques to find better 
agreement between theory and experiments.  

\bigskip 
Using macro-orbital representation of a particle in a many body 
system [15], we developed the long awaited microscopic theory of a 
SIB [16] that explains the properties of liquid $^4He$ with 
unparalleled simplicity, clarity and accuracy.  Clubbing Feynman's 
approach of concluding Eqn. 1 with macro-orbitals, we obtained [16]
$$E(Q) = E(Q)_{mo} = \frac{\hbar^2Q^2}{4mS(Q)} \eqno(2)$$

\noindent
This paper compares Eqn. 2 with $E(Q)_{exp}$ of $He-II$ and 
identifies important aspects of an effective $S(Q)$ to be used 
in this equation.  In addition, it also analyzes the basic 
factors responsible for the experimentally observed anomalous 
nature of small $Q$ phonons of $He-II$ by using another important 
result of our theory ({\it cf.}, Eqn. 5, below).    

\bigskip
\begin{figure}

\includegraphics[angle = -90, width = 0.9\textwidth]{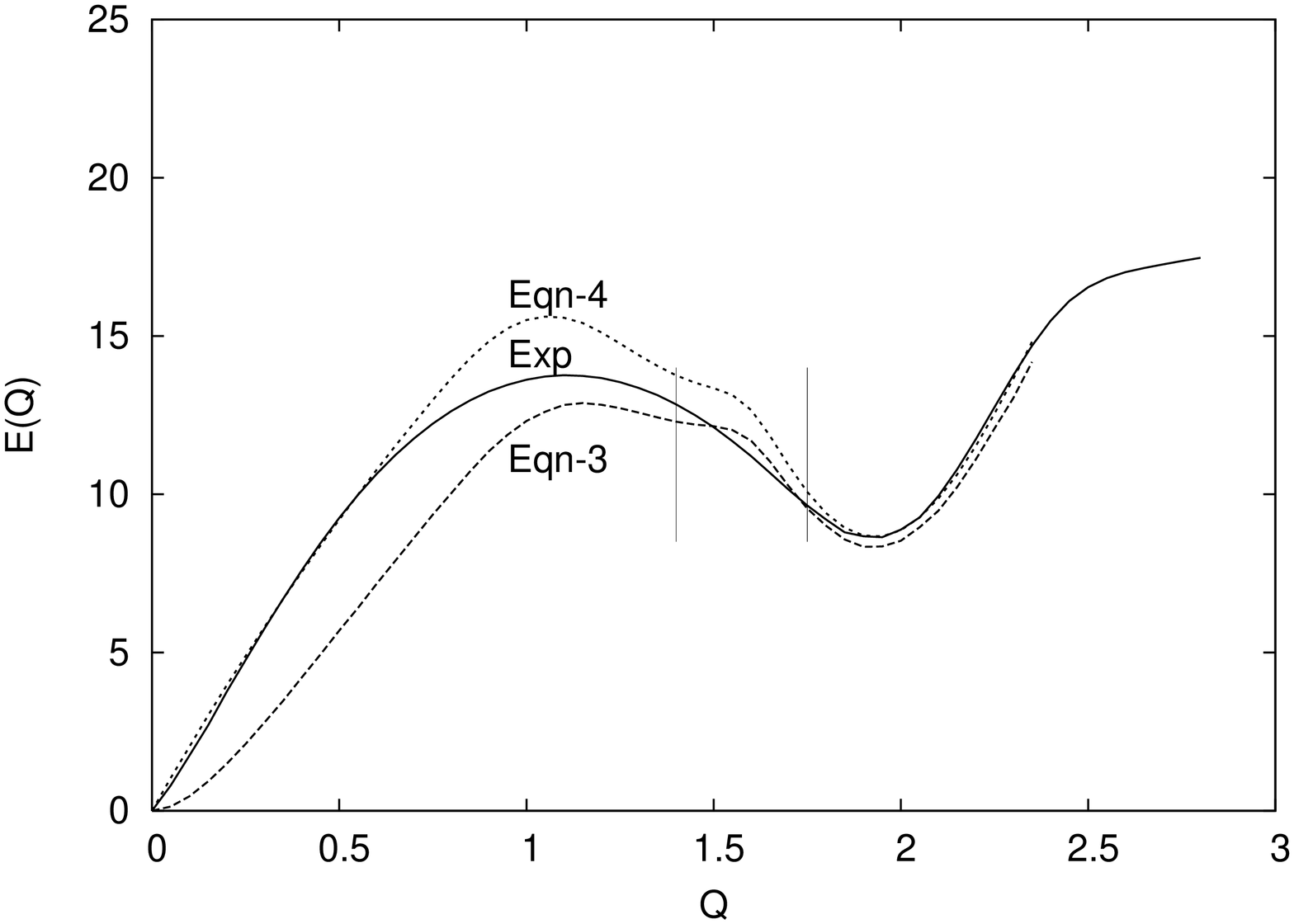}

\bigskip
\noindent
{\small Figure 1. $E(Q)$ (in $^o$K) vs $Q$ (in \AA$^{-1}$) curves 
for $He-II$.  While `Exp' identifies $E(Q)_{exp}$, 
`Eqn-3' and `Eqn-4' represent $E(Q)_{mo}$ and  $E(Q)_{mo}^*$ 
obtained by using Eqns. 3 and 4, respectively.  Both $E(Q)_{mo}$ at $Q$ 
values between two small vertical lines seem to have deviations 
from their normal trend}.  

\end{figure}

\bigskip
\centerline{\bf 2. Elementary Excitations of He-II}

\bigskip
In order to understand the spectrum of elementary excitations of 
$He-II$ in terms of Eqn. 2, we use $S(Q)_{exp}$ and $E(Q)_{exp}$ 
compiled by Donnelly and Barenghi [17].  Appendix I reproduces 
required $S(Q)_{exp}$ and $E(Q)_{exp}$ data for ready reference 
({\it cf.} columns 2 and 5, respectively).  Using $S(Q) = S(Q)_{exp}$ 
in Eqn. 2, we obtain 
$$E(Q)_{mo} = \frac{\hbar^2Q^2}{4mS(Q)_{exp}}  \eqno(3)$$ 

\noindent
which are tabulated in Column 6 of Appendix I and plotted as curve 
`Eqn-3' in Figure 1 where $E(Q)_{exp}$ is also plotted as curve `Exp' 
for comparison.  We find that $E(Q)_{mo}$ does not have (i) desired  
level of agreement with $E(Q)_{exp}$ [the $\Delta E(Q) = E(Q)_{exp}
- E(Q)_{mo}$ is as large as $\approx E(Q)_{mo}$ at 
$Q \approx 0.35$\AA$^{-1}$ ({\it cf.} Appendix I) and the ratio, 
$R = E(Q)_{exp}/E(Q)_{mo}$ tends to have infinitely large value 
when $Q \to 0$], (ii) agreement in the basic nature of the $Q$ dependence 
of $E(Q)_{exp}$ ($\propto Q$) and $E(Q)_{mo}$ ($\propto Q^2$) and 
(iii) consistency with experimentally observed 
{\it non-zero value} ($\approx$ 238.21$m/sec$) of the group velocity 
($V_g)$ and phase velocity ($V_p$) of low $Q$ 
($0$ to $\approx$ 0.5\AA$^{-1}$) phonons in $He-II$.  The following 
analysis concludes that the absence of `(i)' to `(iii)' arises due to 
non-zero value [0.051] of $S(0)$ adopted in [17].    

\bigskip
\begin{figure}
\includegraphics[angle = -90, width = 0.9\textwidth]{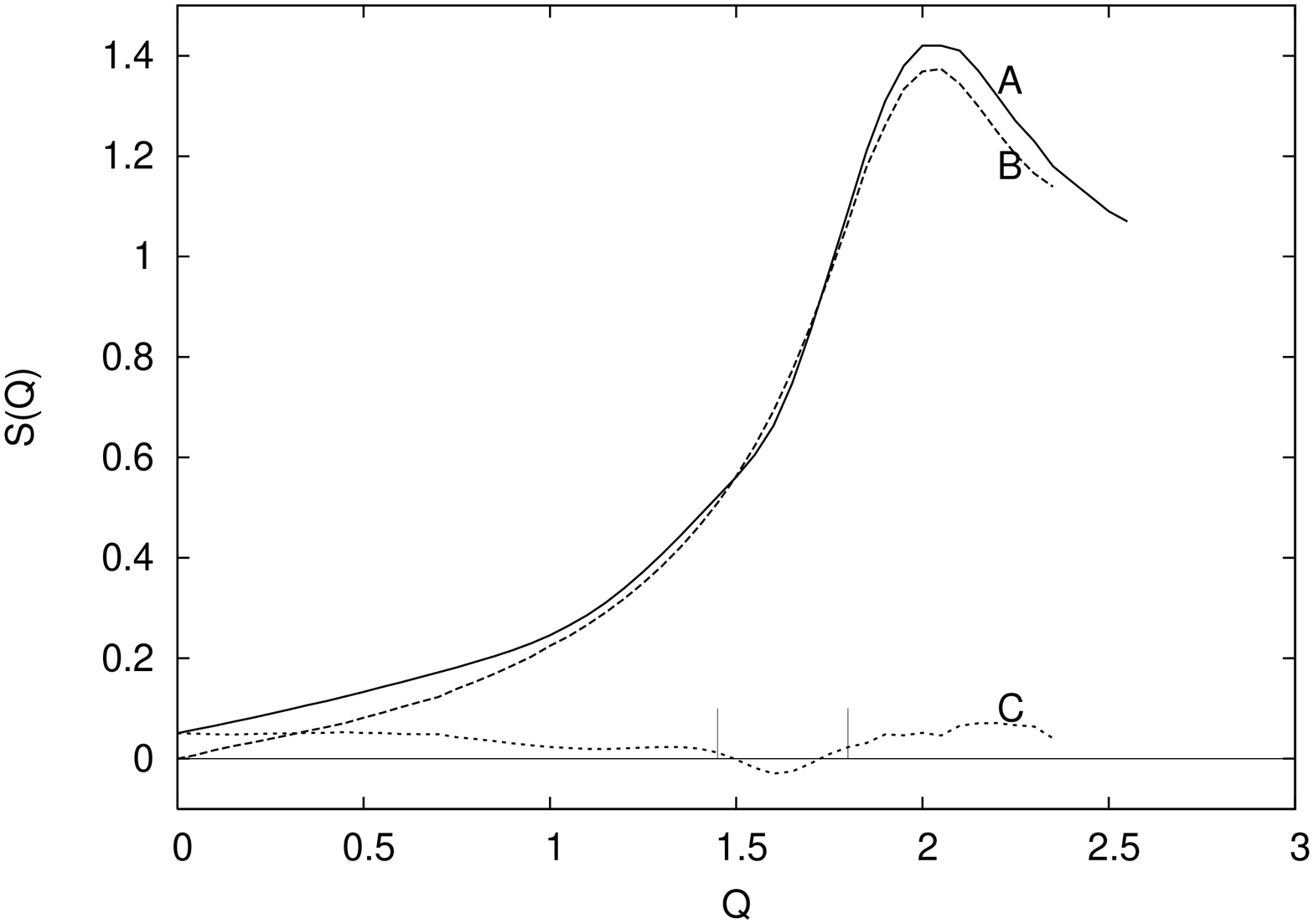}

\noindent
{\small Figure 2.  $S(Q)$ vs $Q$ (in \AA$^{-1}$) curve for liquid 
helium-4.  While Curve-A depicts $S(Q)_{exp}$ (Appendix I, 
column 2), Curve-B represents $S(Q)'$ (Appendix I, column 3) 
which, when used in Eqn. 2, reproduce $E(Q)_{exp}$ (Appendix I, 
column 5) exactly and Curve-C shows $\Delta{S(Q)} = S(Q)_{exp} - 
S(Q)'$ (Appendix I, column 4)).} 

\end{figure}

\bigskip
It is evident that $E(Q)_{exp}$ at low $Q$ is basically a linear 
function of $Q$ which, however, fits with Eqn. 2 only if: (a) 
$S(0)$ is exactly zero, and (b) the major $Q$ dependent term in the 
series expansion of $S(Q)$ at low $Q$ is $\propto Q$.  While 
the $S(Q)_{exp}$ data for $He-II$ ({\it cf.} column 2, Appendix I 
and Figure 2) seem to satisfy `(b)' to a good approximation, but 
$S(0) = 0.051$ contradicts `(a)'.  In case $S(0) \not = 0$, Eqn. 2 
(under the limit $Q \to 0$) renders $E(Q)_{mo}\propto Q^2$ concluding  
$V_g = V_p = 0$ at $Q=0$ which are in total disagreement with 
experiments. Evidently, we need $S(0) = 0$ in place of $S(0) = 0.051$ 
adopted in [17].  In this context we {\it underline} an important 
theoretical inference of Price [18] which states that $S(0)$ 
should be zero for a quantum fluid at {\it zero} temperature 
($T$). While the importance and validity of this inference have been 
emphasized recently by Lamacraft [19], our theoretical inference 
that structural configuration of a quantum fluid at 
$T \le T_{\lambda}$ does not differ from that at $T=0$ [16] implies 
that $S(0)$ should be {\it zero} at all $T$ below $\lambda-$point and 
this is corroborated by the experimental fact that $E(Q)$ of $He-II$ 
hardly depends on $T$.  Evidently, correcting $S(Q)_{exp}$ taken 
from [17], for $S(0)( = 0.051)$, we shift the zero level of $S(Q)$ 
to $S(Q) = 0.051$ and define $S(Q)^* = S(Q)_{exp} - S(0)$ 
as the effective $S(Q)$ that should enter in Eqn. 2 to find  
$$E(Q)_{mo}^* = \frac{\hbar^2Q^2}{4mS(Q)^*} = 
\frac{\hbar^2Q^2}{4m[S(Q)_{exp} - S(0)]} \eqno(4)$$ 

\noindent
as a more appropriate relation to determine $E(Q)$ of a SIB.  
It is obvious that $S(Q)^*$ retains the $Q$ dependence of $S(Q)_{exp}$ 
including its linearity in $Q$ at low $Q$.  We plot $E(Q)_{mo}^*$ in 
Figure 1 ({\it cf.} curve `Eqn-4') for its comparison with $E(Q)_{exp}$.  
We note that the maximum difference in $E(Q)_{mo}^*$ and $E(Q)_{exp}$, 
found to be $\approx 1.8$ at 1.1\AA$^{-1}$, is about 13$\%$ of 
$E(Q)_{exp} \approx 13.8$ which compares with the fractional deviation 
of the recommended values of $E(Q)_{exp}$ and $S(Q)_{exp}$ [17] from 
their adopted database since, as reported in [17], these deviations are 
as large as 3$\%$ for $E(Q)_{exp}$ and 10$\%$ for $S(Q)_{exp}$. 
To this effect we also determined $S(Q)' = \hbar^2Q^2/4mE(Q)_{exp}$ which 
when used in Eqn. 2, obviously reproduce $E(Q)_{Exp}$.  We tabulate $S(Q)'$ 
in Column 3 of Appendix I with its deviation from $S(Q)_{exp}$ [{\it i.e.}, 
$\Delta{S(Q)} = S(Q)_{exp} - S(Q)'$] in Column 4 and plot 
$S(Q)_{exp}$, $S(Q)'$ and $\Delta{S(Q)}$ in Figure 2 as Curves A, B 
and C, respectively.  We note that: (i) $S(Q)_{exp}$ (Curve A) and 
$S(Q)'$ (Curve B) match closely with each other except for some differences 
at low $Q$ which can be attributed to $S(0) = 0.051$ adopted with 
$S(Q)_{exp}$ [17] and (ii) the $\Delta{S(Q)}$ values (Curve C) are positive 
for most $Q$ points (except for points in the small range, $Q=$ 1.5 to 
1.7\AA$^{-1}$, where $\Delta{S(Q)}$ are negative) with average $\bar{\Delta{S(Q)}} 
\approx 0.033$ which justifies our use of $S(Q) = 0.051$ as the zero line of 
$S(Q)^*$.  A careful examination of the recommended $S(Q)_{exp}$ data 
[17] also reveals that the portion of their curve-A (Fig 2) (from $Q = 0$ to 
$Q \approx 1.9$\AA$^{-1}$) has three linear segments: (a) from $Q=0$ to 
$Q \approx 0.9$\AA$^{-1}$, (b) from $Q \approx 0.9$\AA$^{-1}$ to 
$Q \approx 1.5$\AA$^{-1}$, and (c) from $Q \approx 1.5$\AA$^{-1}$ to 
$Q \approx 1.9$\AA$^{-1}$ joined smoothly with each other; this can be 
seen more clearly by plotting $\partial_Q S(Q)_{exp}$ which we have 
not shown here.  This, naturally, implies that $S(Q)_{exp}$ data [17] have 
some systematic errors.  In this context it may be noted that : (i) $E(Q)_{mo}$ 
(curve `Eqn-3', Fig. 1) and $E(Q)_{mo}^*$ (curve `Eqn-4') have deviations from 
their normal trend only in the region from $Q = 1.45$\AA$^{-1}$ to 
$Q = 1.75$\AA$^{-1}$  (identified by two vertical lines in Figs 1 and 2) where 
most $\Delta{S(Q)}$ are negative, and (ii) the $S(Q)_{exp}$ data 
obtained from x-ray diffraction differ significantly from those 
obtained from neutron diffraction [17].  The 
$\Delta{S(Q)}$ data [falling between -0.0293 and 0.0733 (Appendix I) 
around $\bar{\Delta{S(Q)}} \approx 0.033$] also indicate that possible errors of 
$S(Q)_{exp}$ data [17] could be as large as $\bar{\Delta{S(Q)}}$.  Evidently, 
within the limits of such errors of $S(Q)_{exp}$ used here, the differences in our 
theoretical results (Curve `Eqn-4', Fig 1) and experimental results (Curve 
`Exp', Fig 1) are small and this establishes that Eqn. 4 (or Eqn. 2 if 
the available $S(Q)$ data have $S(0) = 0$ level) are accurate enough to provide 
$E(Q)$ of a SIB.      

\bigskip
In the light of what has been inferred from the preceding 
discussion, it is evident that $E(Q)_{Fyn} [= 2E(Q)_{mo}]$ equals 
$\approx 2E(Q)_{exp}$ values not only at high $Q$ but at all $Q$ if 
we enter $S(Q)^*$ for $S(Q)$ in Eqn. 1.  However, if we 
use $S(Q) = S(Q)_{exp}$ as compiled in [17] where $S(0) \not = 0$, 
$E(Q)_{Fyn}$ shows three problems: (i) inconsistency with non-zero 
values of $V_g$ and $V_p$ at $Q=0$ as concluded above for 
$E(Q)_{mo}$ (Eqn.3), (ii) disagreement with $E(Q)_{exp}$ even 
at low $Q$, and (iii) difference in the $Q$ dependence of 
$E(Q)_{exp}$ and $E(Q)_{Fyn}$ at low $Q$.  To this effect we 
determine $E(Q)_{Fyn}$ by using $S(Q)_{exp}$ for $S(Q)$ in Eqn. 1 
and compare it with $E(Q)_{exp}$ at few points of low $Q$ values 
in Table 1.  

\bigskip
\begin{center}
\noindent
Table 1 : Comparison of $E(Q)_{Fyn}$ and $E(Q)_{exp}$
\begin{tabular}{cccc}\hline\hline

$Q$& $\quad$ $E(Q)_{Fyn}$& $\quad$ $E(Q)_{exp}$& $\quad$ 
$R=\frac{E(Q)_{exp}}{E(Q)_{Fyn}}$\\

\hline\hline
0.05& $\quad$ 0.258&$\quad${0.804}&$\quad${$\approx$ 3:1}\\
0.10& $\quad$ 0.918&$\quad${1.747}&$\quad${$\approx$ 2:1}\\
0.15&$\quad$  1.852&$\quad${2.757}&$\quad${$\approx$ 3:2}\\
0.20&$\quad$  2.794&$\quad${3.772}&$\quad${$\approx$ 4:3}\\
0.25&$\quad$  4.222&$\quad${4.772}&$\quad${$\approx$ 6:5}\\
0.30&$\quad$  5.564&$\quad${5.749}&$\quad${$\approx$ 1:1}\\
0.35&$\quad$  6.938&$\quad${6.694}&$\quad${$\approx$ 1:1}\\
0.40&$\quad$  8.430&$\quad${7.598}&$\quad${$\approx$ 1:1}\\
0.50&$\quad$  11.380&$\quad${9249}&$\quad${$\approx$ 5:6}\\

\hline\hline
\end{tabular}
\end{center}

\noindent
Notably the finding, that the difference in the values of 
$E(Q)_{Fyn}$ and $E(Q)_{exp}$ at these $Q$ is not significantly 
large, has been the basis of the inference that $E(Q)_{Fyn}$ 
agrees with $E(Q)_{exp}$ at low $Q$.  However, a careful analysis 
of their relative magnitude ($R = E(Q)_{exp}/E(Q)_{Fyn}$) reveals 
that $R$ at $Q=0.05$\AA$^{-1}$ is as high as {\it three} and it 
is expected to be still higher for $Q < 0.05$ because, to a good 
approximation at low $Q$, $E(Q)_{exp} \propto Q$ and 
$E(Q)_{Fyn} \propto Q^2$ which render $R$ to go as 1/Q with 
$Q \to 0$ and hence implies that $R$ assumes infinitely high value 
at $Q = 0$; in this context it may be noted that $E(Q)_{Fyn}$ 
(Eqn. 1) goes as $Q^2$ at low $Q$ when $S(0) \not = 0$. In 
summary $E(Q)_{Fyn}$ either equals $\approx 2E(Q)_{exp}$ at all 
$Q$ if $S(Q)_{exp}$ data conform to $S(0) = 0$ or its apparent 
agreement with $E(Q)_{exp}$ at low $Q$ as observed by Feynman is 
plagued with above stated three problems.      

\bigskip
\centerline{\bf 3. Anomalous Nature of Phonon Velocities at Low $Q$}

\bigskip
The experimentally observed $V_g$ and $V_p$ of phonons in $He-II$ 
are found to be increasing functions of $Q$ as shown, respectively, 
by Curves E1 and E2 in Figure 3; in this context it may be mentioned 
that experimental $V_g$ are taken from [20], while $V_p = E(Q)/Q$ are 
obtained from experimental $E(Q)$ taken from [17].  It is evident 
that this nature of $V_g$ and $V_p$, limited to small $Q$, is 
opposite to the normal trend of $V_g$ and $V_p$ of phonons in 
crystals.  Various studies of this anomalous character of phonons 
in $He-II$ has been reviewed by Sridhar [21].  In this section 
we examine how this could be understood in the framework of our 
microscopic theory [16].  However, to this effect we do not use 
our $E(Q)_{mo}^*$ data which do not have the required level of 
accuracy due to different possible errors in $S(Q)_{exp}$ used in 
Eqn. 4.  In stead we use another result of our theory which concludes 
that the excitations in $He-II$ type SIB for a large range of $Q 
< Q_{rot}$ (the roton wave vector) can be identified as the 
phonons (longitudinal acoustic modes) of a chain of identical 
atoms (a kind of 1-D crystal) presumed to interact through 
nearest neighbor interaction.  In exact agreement with 
experiments, one expects only a single phonon branch of 
longitudinal modes because the liquid $^4He$ is an {\it isotropic} 
system where atoms experience no {\it shear force} to sustain 
transverse modes. In addition as argued in [16], phonons in this 
chain can have anomalous nature at small $Q$ if its parameters 
({\it viz.} nearest neighbor interaction constant $C$ and 
nearest neighbor distance $d$) in its phonon excitation state 
differ from those in its ground state and this difference 
increases with increasing $Q$.  This can happen the way 
moments of inertia ({\it i.e.}, the structural parameters of 
a molecule) in rotationally excited state differ from those in 
the ground state and the difference increases with increasing 
energy of rotational excitation.  Evidently, the phonon dispersion 
$\omega{(Q)} = E(Q)/\hbar$ of the assumed chain of $^4He$ 
atoms in $He-II$ can be obtained from

\bigskip
\begin{figure}
\includegraphics[angle = -90, width = 0.8\textwidth]{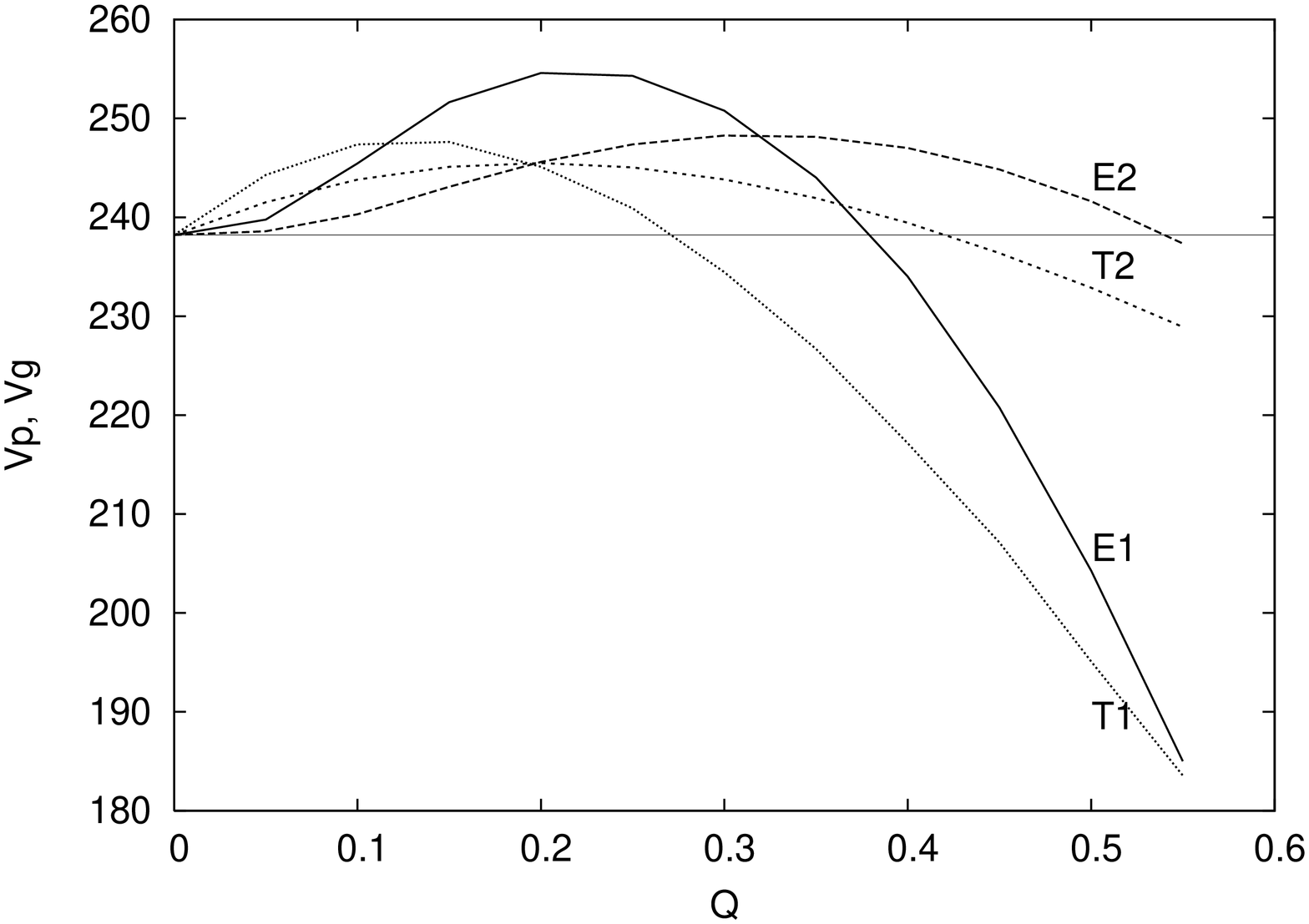}

\noindent
{\small Figure 3. Anomalous variation of $V_g$ and $V_p$ (in m/sec) of 
phonons with $Q$ (in \AA$^{-1}$) for $He-II$.  While 
curves T1 and E1, respectively, represent theoretical and 
experimental $V_g$, curves T2 and E2 are those of $V_p$}.
\end{figure}

$$\omega{(Q)} = \sqrt{\frac{4C(Q)}{m}}|\sin{\frac{Qd(Q)}{2}}| \eqno(5)$$
\noindent
where $C \equiv C(Q)$ and $d \equiv d(Q)$ are assumed to be 
smoothly varying functions of $Q$.  It may be argued that $C(Q)$ 
should increase and $d(Q)$ should decrease with $Q$.  We note that 
the particles, having increased energy under a phonon excitation, 
have smaller quantum size leading them to come closer to each 
other which, obviously, means that $d(Q) \le d(0)$ and $C(Q) \ge 
C(0)$ and they should, respectively, be the decreasing and 
increasing functions of $Q$.  However, since the explicit $Q$ 
dependence of $C(Q)$ and $d(Q)$ is not known we assume 
$$C(Q) = C(0) + C'\sin{\frac{\pi Q}{2Q_{max}}}   
= 3.0554 + 2.5517\sin{\frac{\pi Q}{2Q_{max}}} \eqno (6)$$

\noindent
and 
$$d(Q) = d(0) - d'\sin{\frac{\pi Q}{2Q_{max}}} = 
3.5787 - 0.7484\sin{\frac{\pi Q}{2Q_{max}}}    \eqno (7)$$

\bigskip

\newpage
\begin{center}
\noindent
Table 2 : Theoretical and experimental $V_g$ and $V_p$ in $m/sec^*$; 
correspondence 
with Figure 3 can be set by using T$\equiv theoretical$, 
E$\equiv experimental$, 1$\equiv V_g$, and 2$\equiv V_p$. 

\bigskip
\begin{tabular}{ccccc}\hline\hline

$Q$& $\quad$ $V_g(th)$&$\quad$ $V_g(exp)$&$\quad$ $V_p(th)$&$\quad$ $V_p(exp)$\\

\hline\hline
0.00&$\quad$ 238.21&$\quad$ 238.21&$\quad$ 238.21&$\quad$ 238.21\\
0.05& $\quad$244.29&$\quad$ 239.78&$\quad$ 241.52&$\quad$ 238.59\\
0.10& $\quad$247.37& $\quad$245.44&$\quad$ 243.80&$\quad$ 240.31\\
0.15&$\quad$ 247.62&$\quad$ 251.64&$\quad$ 245.10&$\quad$ 243.08\\
0.20&$\quad$ 245.11&$\quad$ 254.59&$\quad$ 245.49&$\quad$ 245.60\\
0.25&$\quad$ 240.91&$\quad$ 254.31& $\quad$245.04&$\quad$ 247.38\\
0.30&$\quad$ 234.46&$\quad$ 250.78& $\quad$243.83&$\quad$ 248.26\\
0.35&$\quad$ 226.69&$\quad$ 244.02&$\quad$ 241.94&$\quad$ 248.14\\
0.40&$\quad$ 217.16&$\quad$ 234.02&$\quad$ 239.44&$\quad$ 247.01\\
0.45&$\quad$ 207.13&$\quad$ 220.76&$\quad$ 236.39&$\quad$ 244.84\\
0.50&$\quad$ 195.09&$\quad$ 204.27&$\quad$ 232.87&$\quad$ 241.62\\
0.55&$\quad$ 183.56&$\quad$ 185.00&$\quad$ 228.92&$\quad$ 237.36\\

\hline\hline
\end{tabular}
\end{center}
\noindent
The relations (Eqns. 6 and 7) are so assumed to ensure smooth 
variation of $C(Q)$ and $d(Q)$ from $Q=0$ to $Q > Q_{max}$ (maxon 
wave vector) with no discontinuity even in their first $Q$ 
derivative at $Q_{max}$ beyond which they become $Q$ 
independent with $d(Q \ge Q_{max}) = d(Q_{max})$ and 
$C(Q \ge Q_{max}) = C(Q_{max})$.  For $He-II$, we fix $C(0)$, 
$C'$, $d(0)$ and $d'$ empirically by using certain experimental 
data.  For example $d(0) = 3.5787\AA$ is fixed by using 
$He-II$ density = $0.1450 gm/cc$ [22], while fixation of $C(0) = 3.0554 
dyne/cm$ uses $Q=0$ phonon velocity = 238.21 $m/sec$.  
Similarly, the smallest $d [= d(0) - d']$ is equated to the 
hard core size ($\sigma$) of particles since two particles can 
not have $d < \sigma$.  We obtained $\sigma = 2.8293\AA$  for 
$^4He$ atoms by assuming that phonon energy for the chain having 
$d = \sigma$ should have its maximum at $Q=Q_{max}$ which equals 
1.11 \AA$^{-1}$ for $He-II$ and this means that $\sigma = 
\pi/1.11$\AA$^{-1}$ and $d' = 0.7484$\AA.  Likewise we use 
$E(Q_{max}) = 13.92 K$ [8] and $d(Q_{max}) = \sigma$ to fix 
highest $C(Q) [= C(0) + C'] = 5.6071 dyne/cm$ and $C' = 
3.5517 dyne/cm$.  This concludes that Eqns. 6 and 7 can help 
in determining $C(Q)$ and $d(Q)$ to obtain $E(Q)$ at any $Q$ 
ranging from $Q=0$ to $Q = Q_{max}$.  To obtain $E(Q > Q_{max})$, 
it is assumed that the chain parameters for $Q \ge Q_{max}$ do 
not differ from $C(Q_{max}) = 5.6071$  and $d = \sigma$.  Using 
Eqns. 6 and 7 in  Eqn. 5, we calculated $E(Q)$, $V_g$ and $V_p$.  
While $E(Q)$ so obtained is tabulated as $E(Q)_{eq5}$ in column 8 
of Appendix I, $V_g$ and $V_p$ are tabulated in Table 2 ({\it cf.}, 
columns 2 and 4, respectively).  We plot 
our theoretical $V_g$ and $V_p$ in Figure 3 ({\it cf.} Curves 
T1 and T2, respectively) along with their experimental values 
({\it cf.} Curves E1 and E2, respectively).  It is important 
to note that our theoretical $V_g$ and $V_p$ exhibit anomalous 
nature of phonons as shown by experiments and their values 
match closely with experimental values.  While a maximum 
quantitative difference of $\approx 5\%$ for $V_g$ and 
$\approx 3\%$ for $V_p$ is not very significant but a 
qualitative difference in the nature of their variation 
near $Q \approx 0$ indicates that the present choice of the $Q$ 
dependence of $C(Q)$ and $d(Q)$ (Eqns. 6 and 7) need to be replaced 
by a better choice.  We hope to find the desired choice in our 
future course of studies.  

\bigskip
\begin{figure}
\includegraphics[angle = -90, width = 0.8\textwidth]{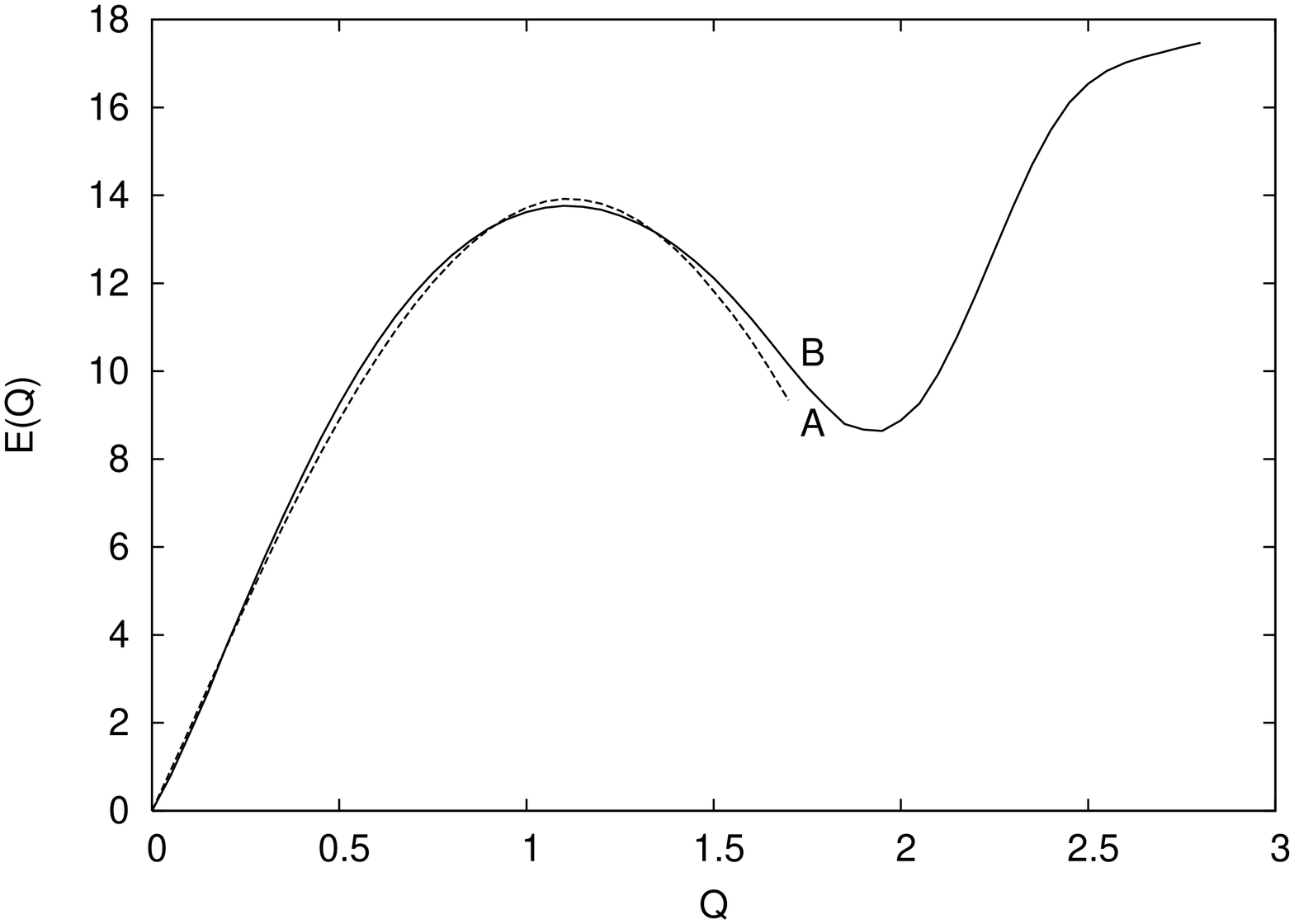}

\noindent
{\small Figure 4. $E(Q)$ of $He-II$ : (Curve-A) $E(Q)_{eq5}$ represents 
the energy dispersion of phonons in a chain of identical atoms 
whose parameters change with $Q$ and (Curve-B) $E(Q)_{exp}$ of 
phonons obtained from [17]}. 
\end{figure}

\bigskip
We plot our theoretical $E(Q)_{eq5}$ ({\it cf. } column 8 of 
Appendix I), obtained from Eqn. 5, in Figure 4 (Curve A) along with a 
plot of $E(Q)_{exp}$ (Curve B).  The fact that these two $E(Q)$ 
match very closely for all $Q$ ranging from $Q = 0$ to  $Q$ close 
to $Q_{rot}$ implies that $E(Q)_{exp}$ of $He-II$ over this range 
can, really, be accounted for by the phonons of a chain of identical 
atoms clubbed with what we may call as 'final state effect'.  
Interestingly, recasting Eqn. 4 as 
$$E(Q)_{mo} = \frac{\hbar^2Q^2}{2m^*}  \eqno(8)$$ 

\noindent 
one may identify that the excitations of a SIB like liquid $^4He$ 
are nothing but the motions of a single particle whose mass changes 
with $Q$ as $m^* \approx 2m[S(Q)_{exp} - S(Q=0)]$ which can be known 
as effective mass.  We note that $m^*$ for $He-II$ changes from 
$m^* = 0$ at $Q=0$ to 
$m^* \approx m$ for $Q >> Q_p$ (the lowest $Q$ of plateau modes) 
through its maximum value of $\approx 2.74m$ for $Q \approx Q_{rot}$ 
which falls close to the $Q$ of the peak in $S(Q)$.  However, in the 
light an excellent agreement between $E(Q)_{eq5}$ and $E(Q)_{exp}$ 
(Figure 4), it will be more appropriate to state that excitations 
in liquid $^4He$ are : (i) phonon like for $Q \le 2\pi/d$, 
(ii) phonon dressed single particle motions for $2\pi/d \le Q \le 
Q_p$, and (iii) an almost free single particle motion 
for $Q > Q_p$; note that $Q = 2\pi/d$ and $Q = Q_p$ 
should not be considered as sharply dividing $Q$ points of 
the three ranges.  
It may be argued that this shift in the nature of the excitations  
arises from the facts that: (i) particles in liquid $^4He$ type 
SIB do not have fixed locations like those in solids; in spite 
of their nearly localized positions, they remain free to move in 
order of their locations, (ii) the excitation wave length 
$\lambda_{ex}$ for $Q > 2\pi/d$ becomes shorter than $d$ implying 
that the energy and momentum of the excitation can be carried by 
a single particle, and (iii) a particle of $\lambda_{ex} < \sigma$ 
({\it i.e.}, $Q > 2\pi/\sigma$), can be expected to behave almost 
like a free particle since it does not have any wave superposition 
even with its nearest neighbor.      

\bigskip
\centerline{\bf 4. Concluding Remarks}

\bigskip
The energy of elementary thermal excitations of liquid helium-4 
obtained by using $E(Q) = \hbar^2Q^2/4mS(Q)$ (concluded from our 
recent microscopic theory of a SIB [16]) matches closely with 
$E(Q)_{exp}$ of $He-II$ when $S(Q)$ data set used for this 
purpose conforms to $S(0) = 0$ which is concluded to be an 
important aspect of a superfluid at $T=0$ [18]; interestingly, 
an important conclusion of our theory [16] that the relative 
configuration of particles in the superfluid state of a SIB is 
independent of $T$ implies that $S(0) = 0$ should be valid for 
all $T \le T_{\lambda}$ and this is corroborated by the fact 
that $E(Q)_{exp}$ of $He-II$ hardly change with $T$.  Our theory 
also explains the anomalous nature of low $Q$ phonons in $He-II$.  
Since the excitation spectrum of Bose condensed dilute gases too 
is found to be qualitatively identical to that of $He-II$ [23], 
our theory can similarly explain the related properties of BEC 
state of such gases.  We note that $He-I$ and $He-II$ as per our theory 
differ from each other in relation to the relative configuration 
of their particles whose details are discussed in [16].  However, 
the important difference of $He-I$ and $He-II$ which affects 
$E(Q)$ can be identified with the collisional motion of particles 
in $He-I$ and collision-less state of particles in $He-II$ [16].  
Consequently, $E(Q)$ of the two phases is expected to 
differ significantly not in the positions of their excitations 
but in their line-widths which should be much larger for $He-I$ 
than for $He-II$ and this agrees with experiments.  
In addition this study shows that: (i) $E(Q)_{exp}$ 
values of $He-II$ are consistent with $S(0)= 0$ (not with 
$S(0) \not = 0$) and (ii) $E(Q)_{Fyn}$ (Eqn. 1) either equals 
$\approx 2E(Q)_{exp}$ at all $Q$ if $S(Q)$ data conform to 
$S(0)= 0$ or its apparent agreement with $E(Q)_{exp}$ at low $Q$ 
seen for $S(0) \not = 0$ is plagued with {\it three problems} 
concluded in Section 2.   As such all these facts provide good 
evidence of the accuracy of Eqn. 4 which implies that our theory 
provides strong foundation to understand the thermodynamic 
properties of a SIB.  We would use our theory for a similar 
analysis of the hydrodynamic behavior of a SIB in our future 
course of studies.  It should be interesting to note that our theory, 
for the first time, provided 
highly accurate account of several properties of $He-II$, {\it e.g.} 
: (i) logarithmic singularity of specific heat of $He-II$ and related 
properties, (ii) quantized vortices with an appropriate answer to the 
question raised by Wilks [22] on the validity of Feynman's explanation 
for their origin, (iii) $T^3$ dependence of specific heat, (iv) 
two fluid behavior with a conclusion that each particle 
simultaneously participates in superfluid and normal components, (v) 
superfluidity of quasi 1-D and 2-D systems.

\newpage

\centerline{\bf Appendix I}

\bigskip

\begin{center}

Data pertaining to a comparative study of $E(Q)_{mo}$ 
and $E(Q)_{exp}$. The $Q$ is in \AA$^{-1}$ and $E(Q)$ in $^o$K 

\begin{tabular}{cccccccc}\hline\hline

$Q$& $S(Q)_{exp}$& $S(Q)'$& $\Delta{S(Q)}$& $E(Q)_{exp}$& $E(Q)_{mo}$& $E(Q)_{mo}^*$& $E(Q)_{eq5}$\\\hline\hline
0.00& 0.0510& 0.0000& 0.0510& 0.000& 0.000& 0.000& 0.00\\
0.05& 0.0583& 0.0076& 0.0500& 0.804& 0.129& 1.037& 0.92\\
0.10& 0.0659& 0.0173& 0.0486& 1.747& 0.459& 2.032& 1.86\\
0.15& 0.0736& 0.0252& 0.0484& 2.757& 0.926& 3.015& 2.81\\
0.20& 0.0815& 0.0321& 0.0494& 3.772& 1.487& 3.971& 3.75\\
0.25& 0.0897& 0.0397& 0.0500& 4.772& 2.111& 4.891& 4.68\\
0.30& 0.0980& 0.0474& 0.0506& 5.749& 2.782& 5.799& 5.59\\
0.35& 0.1065& 0.0554& 0.0511& 6.694& 3.469& 6.684& 6.47\\
0.40& 0.1152& 0.0632& 0.0520& 7.598& 4.215& 7.547& 7.32\\
0.45& 0.1242& 0.0710& 0.0532& 8.452& 4.940& 8.378& 8.13\\
0.50& 0.1333& 0.0818& 0.0515& 9.249& 5.690& 9.199& 8.89\\
0.55& 0.1427& 0.0918& 0.0509& 9.979& 6.409& 9.990& 9.62\\
0.60& 0.1522& 0.1025& 0.0497& 10.641& 7.170& 10.773& 10.29\\
0.65& 0.1620& 0.1139& 0.0481& 11.237& 7.900& 11.527& 10.92\\
0.70& 0.1720& 0.1229& 0.0491& 11.767& 8.630& 12.264& 11.50\\
0.75& 0.1822& 0.1393& 0.0429& 12.232& 9.360& 12.984& 12.02\\
0.80& 0.1928& 0.1535& 0.0393& 12.633& 10.040& 13.668& 12.48\\
0.85& 0.2040& 0.1688& 0.0352& 12.971& 10.730& 14.301& 12.89\\
0.90& 0.2164& 0.1852& 0.0312& 13.248& 11.362& 14.831& 13.23\\
0.95& 0.2303& 0.2032& 0.0271& 13.464& 11.880& 15.243& 13.51\\
1.00& 0.2463& 0.2225& 0.0238& 13.620& 12.316& 15.506& 13.72\\
1.05& 0.2648& 0.2435& 0.0213& 13.717& 12.606& 15.616& 13.86\\
1.10& 0.2862& 0.2664& 0.0198& 13.757& 12.820& 15.580& 13.92\\
1.15& 0.3109& 0.2916& 0.0193& 13.740& 12.880& 15.410& 13.90\\
1.20& 0.3395& 0.3192& 0.0203& 13.667& 12.830& 15.116& 13.81\\
1.25& 0.3716& 0.3496& 0.0220& 13.540& 12.720& 14.759& 13.65\\
1.30& 0.4066& 0.3833& 0.0233& 13.359& 12.580& 14.393& 13.42\\
1.35& 0.4438& 0.4206& 0.0232& 13.125& 12.430& 14.051& 13.12\\
1.40& 0.4825& 0.4625& 0.0200& 12.839& 12.290& 13.756& 12.76\\
1.45& 0.5219& 0.5096& 0.0123& 12.503& 12.204& 13.521& 12.33\\
1.50& 0.5614& 0.5625&-0.0011& 12.118& 12.150& 13.350& 11.85\\
1.55& 0.6049& 0.6232&-0.0183& 11.684& 12.030& 13.135& 11.30\\
1.60& 0.6633& 0.6926&-0.0293& 11.202& 11.690& 12.662& 10.70\\
1.65& 0.7466& 0.7724&-0.0258& 10.680& 11.040& 11.852& 10.04\\
1.70& 0.8529& 0.8623&-0.0094& 10.148& 10.260& 10.914& 9.34\\

\hline\hline 

\end{tabular}

\end{center}

\newpage
\noindent
Appendix I contd.

\begin{center}
\begin{tabular}{cccccccc}\hline\hline

$Q$& $S(Q)_{exp}$& $S(Q)'$& $\Delta{S(Q)}$& $E(Q)_{exp}$& $E(Q)_{mo}$& $E(Q)_{mo}^*$& $E(Q)_{eq5}$\\\hline\hline

1.75& 0.9721& 0.9626& 0.0095& 9.644& 10.069& 9.706\\
1.80& 1.0940& 1.0671& 0.0269& 9.204& 9.408& 9.132\\
1.85& 1.2100& 1.1784& 0.0316& 8.866& 8.942& 8.693\\
1.90& 1.3080& 1.2616& 0.0464& 8.667& 8.697& 8.433\\
1.95& 1.3800& 1.3334& 0.0466& 8.644& 8.664& 8.421\\
2.00& 1.4180& 1.3684& 0.0496& 8.833& 8.861& 8.601\\
2.05& 1.4240& 1.3736& 0.0504& 9.271& 9.269& 9.030\\
2.10& 1.4050& 1.3443& 0.0607& 9.941& 9.863& 9.337\\
2.15& 1.3690& 1.2992& 0.0698& 10.784& 10.621& 10.290\\
2.20& 1.3230& 1.2497& 0.0733& 11.742& 11.523& 11.168\\
2.25& 1.2730& 1.2031& 0.0699& 12.751& 12.546& 12.135\\
2.30& 1.2260& 1.1657& 0.0603& 13.753& 13.634& 13.084\\
2.35& 1.1800& 1.1391& 0.0409& 14.687& 14.813& 14.228\\

\hline\hline 

\end{tabular}

\end{center}


\bigskip

\begin{thebibliography}{??}
\bibitem[1]{Kn:gnus}
H. Glyde, Excitations in Liquid and Solid Helium, 
Oxford University Press (2002).
\bibitem[2]{Kn:gnus}
D. Henshaw and A.D.B. Woods, Phys. Rev. {\bf 121}, 1266 (1961).
\bibitem[3]{Kn:gnus}
R.A. Cowley and A.D.B Woods, Phys. Rev. Lett.{\bf 21}, 787 (1968)
\bibitem[4]{Kn:gnus}
R.A. Cowley and A.D.B Woods, Can J. Phys. {\bf 49}, 177 (1971).
\bibitem[5]{Kn:gnus}
O.W. Dietrich, E.H. Graf, C.H. Huang and L. Passell, Phys. Rev. 
{\bf A5}, 1377 (1972).
\bibitem[6]{Kn:gnus}
E.C. Svensson, A.D.B. Woods and P. Martel, Phys. Rev. Lett. 
{\bf 29} 1148 (1972).
\bibitem[7]{Kn:gnus}
F. Mezei, Phys. Rev. Lett. {\bf 44}, 1601 (1980).
\bibitem[8]{Kn:gnus}
A.D.B. Woods and R.A. Cowley, Rep. Prog Phys. {\bf 36}, 
1135 (1973).  
\bibitem[9]{Kn:gnus}
H.R. Glyde and E.C. Svensson in {\it Methods of Experimental Physics}, 
Eds. D.L. Price and K. Skold, Academic Press. Inc. London (1987) 
Vol 23 Part B p 303. 
\bibitem[10]{Kn:gnus}
(a) R.P. Feynman, Phys. Rev. {\bf 94}, 262 (1954).

(b) R.P. Feynman, in {\it Progress in Low Temperature Physics},
edited by C.J. Groter, (North Holland, Amsterdem, 1955), 
{\bf Vol. 1}, p. 17. 

(c) R.P. Feynman in {\it Statistical Mechanics}, (Benjamin, 1976), p. 284.

\bibitem[11]{Kn:gnus}  
R.P. Feynman and M. Cohen, Phys. Rev. {\bf 102}, 1189(1956)
\bibitem[12]{Kn:gnus}
K E. Schmidt and V.R. Pandharipande, 
Phys. Rev. {\bf B21}, 3945(1980)
\bibitem[13]{Kn:gnus}
E. Manousakis and V.R. Pandharipande, 
Phys. Rev. {\bf B30}, 5062(1984)
\bibitem[14]{Kn:gnus}
M. Saalera, Phys. Rev. {\bf B33}, 4596(1986).
\bibitem[15]{Kn:gnus}
(a) Y.S. Jain, {\it Basic Foundations of the Microscopic Theory of 
Superconductivity} arxiv.org/Cond-mat/0603784.

(b) Y.S. Jain, {\it Ground state of a System of N Hard Core Quantum 
Particles} arxiv.org/Cond-mat/0606409.

(c) Y.S. Jain, {\it Wave Mechanics of Two Hard Core Quantum Particles 
in a 1-D Box} arxiv.org/quant-ph/0603233, 
Central Euro. J. Phys. {\bf 2}, 709 (2004).

(d) Y.S. Jain, {\it Unification of the Physics of Bosonic and Fermionic 
Systems}, Ind. J. Phys. {\bf 79} 1009 (2005).  

\bibitem[16]{Kn:gnus}
Y.S. Jain, {\it Macro-Orbitals and Microscopic Theory of a System of 
Interacting Bosons} arxiv.org/Cond-mat/0606571.
\bibitem[17]{Kn:gnus}
R.J. Donnelly and C.F. Barenghi, {\it The Observed Properties of 
Liquid Helium at Saturated Vapor Pressure } http://darkwing.uoregon.
edu/~rjd/vapor1.htm, Chap 14.
\bibitem[18]{Kn:gnus}
P.J. Price, Phys. Rev. {\bf 94}, 257 (1954) 
\bibitem[19]{Kn:gnus}
A. Lamacraft, {\it Particle correlations in a fermi superfluid}, 
 arxiv.org/Cond-mat/0510111. 
\bibitem[20]{Kn:gnus}
R.J. Donnely, J.A. Donnelly and R.N. Hill,
J. Low Temp. Phys. {\bf 44} 471 (1981).
\bibitem[21]{Kn:gnus}
R. Sridhar, Phys. Rep. {\bf 146}, 249 (1986).
\bibitem[22]{Kn:gnus}
J. Wilks, {\it The Properties of Liquid and Solid Helium}, 
Clarendon Press, Oxford (1967) p. 666.
\bibitem[23]{Kn:gnus}
D.H.J. O'Dell, S. Giovannazzi and G. Kurizki, Phys. Rev. Lett. 
{\bf 90}, 110402 (2003). 
 
\end{thebibliography}
\end{document}